\documentclass[12pt]{iopart}
\usepackage{graphicx}
\usepackage{xcolor}
\usepackage{amssymb}
\usepackage{siunitx}
\usepackage{lineno}
\usepackage{xspace}
\usepackage{subcaption}

\usepackage[backend=biber,style=phys,sorting=none]{biblatex} 

\usepackage{hyperref}
\hypersetup{linktocpage} 
\usepackage{enumerate} 

\newcommand{\FP}{Fabry-Perot\xspace}

\begin{document}

\title{Toward Low-Latency, High-Fidelity Calibration of the LIGO Detectors with Enhanced Monitoring Tools}

\author{M. Wade$^1$, J. Betzwieser$^2$, D. Bhattacharjee$^1$, L. Dartez$^2$, E. Goetz$^3$, J. Kissel$^4$, L. Sun$^5$, A. Viets$^6$, M. Carney$^1$, E. Makelele$^1$, L. Wade$^1$}
\address{$^1$Kenyon College, Gambier, Ohio}
\address{$^2$LIGO Livingston Observatory, Livingston, Louisiana}
\address{$^3$University of British Columbia, Vancouver, BC V6T 1Z4, Canada}
\address{$^4$LIGO Hanford Observatory, Richland, Washington}
\address{$^5$OzGrav-ANU, Centre for Gravitational Astrophysics, Research School of Physics and Research School of Astronomy \& Astrophysics, The Australian National University, Canberra, Australian Capital Territory 2601, Australia}
\address{$^6$Concordia University Wisconsin, Mequon, Wisconsin}
\ead{wadem@kenyon.edu}
\vspace{10pt}

\begin{abstract}
Accurate and reliable calibration of the Advanced LIGO detectors has enabled a plethora of gravitational-wave discoveries in the detectors' first decade of operation, starting with the ground-breaking discovery, GW150914.  In the first decade of operation, the calibrated strain data from Advanced LIGO detectors has become available at a lower latency and with more reliability.  In this paper, we discuss the relevant history of Advanced LIGO calibration and introduce new tools that have been developed to enable faster and more robust calibrated strain data products in the fourth observing run (O4).  
We discuss improvements to the robustness, reliability, and accuracy of the low-latency calibration pipeline as well as the development of a new tool for monitoring the LIGO detector calibration in real time.
\end{abstract}

%
%
%
%
%


\section{Introduction}
\label{sec:introduction}
The direct detection of gravitational waves (GWs) by the Advanced Laser Interferometer Gravitational-wave Observatory (LIGO)~\cite{Aasi2015aLIGO} detector in 2015 marked the beginning of a new era in observational astrophysics and fundamental physics. The first GW detection, GW150914~\cite{abbott2016gw150914main,abbott2016gw150914cbcsearch}, resulted from the merger of a binary black hole system, confirming a key prediction of general relativity and demonstrating the feasibility of ground-based interferometric GW detection. 
Since this landmark event, the field has experienced rapid progress, with hundreds of detections of binary black hole and neutron star mergers by the global GW detector network of LIGO, Virgo~\cite{acernese2014}, and KAGRA~\cite{akutsu2020}.

In the first three observing runs, O1--O3, in total 90 confident detections were made of GWs from compact binary mergers~\cite{Abbott2023gwtc3}.
The fourth observing run, O4, is ongoing at the writing of this paper with a nearly doubled detection rate and significant improvement in detector performance~\cite{detection-rate, capote2025advanced}. After O4, major upgrades are planned in LIGO, Virgo, and KAGRA, with the aim of increasing the sensitivity, as measured by the distance to which a binary neutron star merger event would be seen at a fiducial signal-to-noise ratio, by more than a factor of two. Thus, in the upcoming fifth observing run, O5~\cite{obs-scenario}, we expect an order-of-magnitude increase in the number of detections, requiring low-latency and high-quality data products to facilitate rapid astrophysical discoveries.  

A fundamental component of the success of GW observations is the precise calibration of interferometric data, which reconstructs the raw, digitized electrical output of the detector into an accurate and reliable measure of the dimensionless strain~\cite{abbott2017calibration,cahillane2017calibration,sun2020characterization}. Calibration accuracy directly impacts source parameter estimation, waveform reconstruction, and tests of general relativity, making it a critical aspect of GW astronomy~\cite{Schutz2018,hall2019systematic}. Miscalibrations can introduce systematic biases, affecting the astrophysical interpretation of detected signals, limiting the reliability of inferred source properties, and leading to false deviations from general relativity~\cite{Gupta2025}.

\begin{figure}
    \centering
    \includegraphics[width=0.75\linewidth]{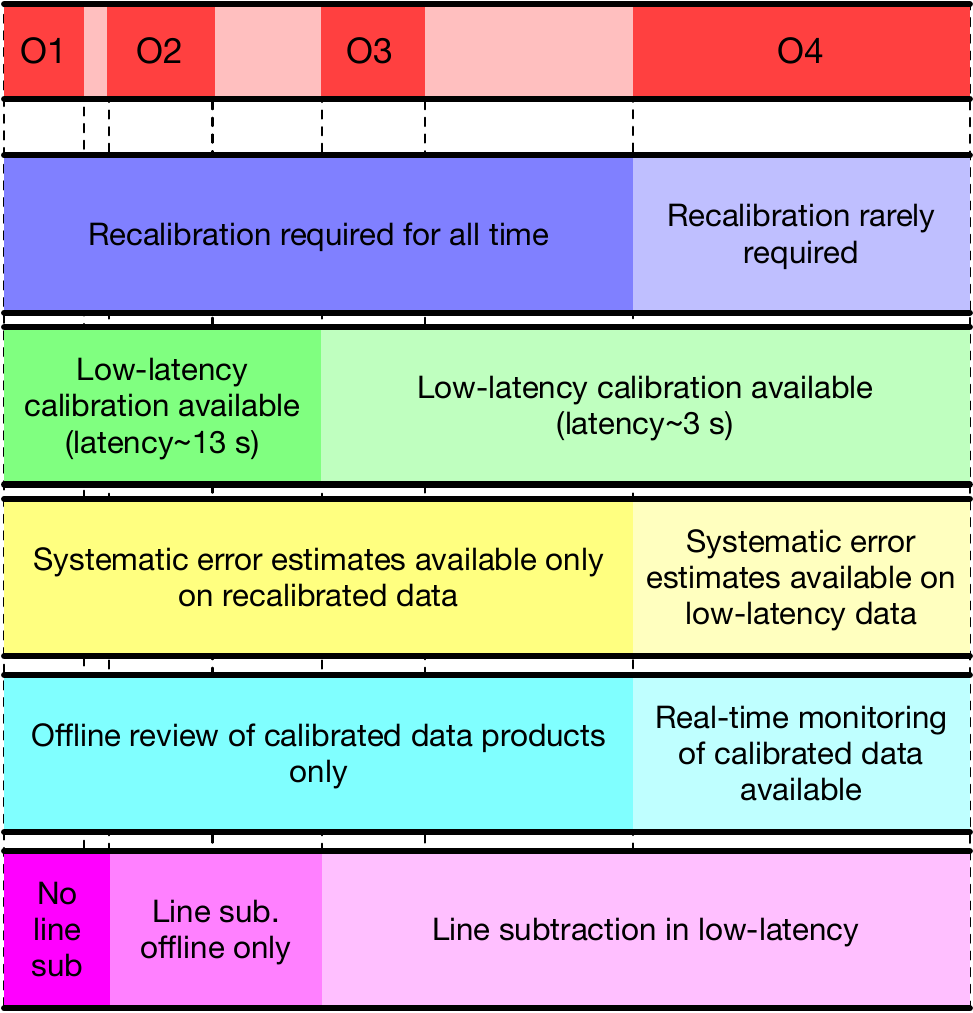}
    \caption{Schematic visualizing the evolution of several aspects of LIGO calibration from O1 through O4.  This graphic focuses on five specific components of the LIGO calibration procedure which are discussed primarily in section~\ref{sec:currentwork}.}
    \label{fig:evolution}
\end{figure}

At the time of the detection of GW150914~\cite{abbott2017calibration} and throughout O1--O3, calibration was performed with high accuracy but at high latency, with the most precise strain data becoming available only after careful post-processing with a significant delay on the order of several months. 
In contrast, calibration methodologies implemented in O4 enable the real-time production of publishable strain data along with a well-quantified systematic error estimate on this strain data.
Figure~\ref{fig:evolution} summarizes the evolution of various features in the LIGO calibration process from O1 through O4, which are discussed in more detail in section~\ref{sec:currentwork}.
With further sensitivity improvements in GW observatories, which allow for high signal-to-noise ratio observations, the next major challenge is the development of calibration techniques that achieve real-time calibration with an even higher accuracy, ensuring that precision science can be performed immediately upon detection.
This advancement will be crucial for multi-messenger astronomy and for rapid source parameter estimation with high fidelity in future observing runs.

Although LIGO, Virgo, and KAGRA share similar core calibration schemes~\cite{Acernese_2022,akutsu2020}, each has developed distinct calibration pipelines and adopts different combinations of absolute references~\cite{aubin2024,estevez2018,estevez2021,Inoue2018}. This paper focuses on the methods and developments specific to LIGO.
The paper is organized as follows: \sref{sec:background} provides a robust overview of the historical background of calibration in Advanced LIGO over the decade following the discovery of GW150914, \sref{sec:cal_challenges} discusses the expected challenges and calibration goals in the next observing run (O5), \sref{sec:currentwork} describes new techniques and tools that have been implemented to improve the accuracy and latency of calibrated LIGO data, \sref{sec:nextgen} discusses calibration requirements in the next generation GW detectors, and we conclude in \sref{sec:conclusion}.

\section{Background and Historical Overview of LIGO Calibration}
\label{sec:background}
\label{sec:background}

    In order to maintain the interferometer at its operating point for extended periods of time, the LIGO detectors depend on a multitude of control loops that actuate on various systems in response to sensor signals in real time. 
    The primary control loop of astrophysical interest interferometrically maintains the difference in \FP arm cavity lengths. 
     This differential arm, or DARM, degree of freedom is most sensitive to gravitational waves, and the control loop responsible for the DARM degree of freedom is referred to as the DARM loop.
    The DARM loop sensor is the photodetector that measures laser light at the output of the interferometer (anti-symmetric side of the beamsplitter) and the actuators are the quadruple pendulum actuators that move the test masses to keep the \FP cavities on resonance.
    Displacements in differential arm cavity length, whether from internal (e.g. interferometer actuators) or external (e.g. gravitational waves or noise) sources are encoded in the DARM loop error signal.

    Calibration is the process of reconstructing the DARM displacement, $\Delta L_{\textrm{free}}$, 
   from the DARM loop error signal, $d_{\textrm{err}}$. 
    In the frequency domain, the DARM loop error signal and the DARM displacement are related by
    \begin{equation}
    \label{eq:deltal_response}
        \tilde d_{\rm err}(f) = \frac{\Delta \tilde L_{\textrm{free}}(f)}{R(f)} ~,
    \end{equation}
    where $R(f)$ is the frequency-dependent DARM loop response (also commonly referred to as the ``detector response'') and throughout the paper a tilde refers to the Fourier transform of a time-domain signal.

While $d_{\textrm{err}}$ is measured continually, $R$ is only fully measured infrequently during breaks in gravitational-wave data collection. 
A model of $R$ is constructed during these breaks and is used to extrapolate to other times.
Therefore, the accuracy of the calibration is dependent on the accuracy of the modeled detector response $R$.  
That model, maintained by the LIGO calibration team, is validated in the frequency band in which the detector is most sensitive to gravitational waves (10~Hz to 5,000~Hz). 

During observing runs, the detector response typically does not change significantly, but small changes can impact the overall systematic error and uncertainty in the reconstructed $\Delta L_{\rm free}$ signal and therefore must be closely monitored.
The detector response varies on time scales longer than one second, due to, for example, thermal effects, alignment fluctuations, and seismic activity. 
The time-dependence of the response is tracked by constant measurements of the response using fiducial displacements (see \sref{sec:dispreferences}) at strategically selected frequencies in order to minimize any disturbances to the collection of astrophysical signals. 
The model of the detector response is then reconstituted, taking the tracked temporal changes into account.  The final reconstructed $\Delta L_{\rm free}$ uses this reconstituted detector response in order to achieve the highest level of accuracy on the calibrated data.
The effort to produce an accurate calibration comprises maintaining absolute reference standards, taking frequent measurements of the individual DARM loop components, constructing and validating a reliable model of the detector response, deploying an application of \eref{eq:deltal_response} to reconstruct $\Delta L_{\rm free}$, and estimating the systematic error and uncertainty associated with the detector response model.

    \subsection{Absolute References}
    \label{sec:absref}
    The fundamental quantities that define the response of the detector are the amount of power circulating in the \FP cavities, the mass of the cavity mirrors (including that of the suspension that supports them), and the frequency tracking and/or timing system. 
    The technical quantities that are critical for characterizing the control system surrounding that response are electronic, defined by functions converting current to voltage, raw to conditioned voltage, or voltage to current. 
    These fundamental and technical quantities must be measured with high accuracy and precision in order to limit the amount of systematic error and uncertainty in the calibrated detector output. 
    
    Searches for gravitational waves are only quadratically sensitive to detector calibration systematic error or uncertainty~\cite{allen1996ligo,lindblom2008model,brady2008interpreting}.
    Thus, collective systematic error and uncertainty in the calibrated data below 10\% in magnitude and 10~deg in phase, independent of frequency and time has not limited \emph{detection}. 
    However, once detected, the process of estimating the astrophysical source properties of the gravitational wave is linear in detector calibration uncertainty and systematic error~\cite{payne2020gravitational,vitale2021physical,essick2022calibration,huang2025impact,yousuf2023effects,Sinha2025}. 
    Additionally, calibration uncertainty has been shown to always reduce the sensitivity of modeled searches for gravitational waves \cite{PhysRevD.105.082002}.
    As such, we now strive for a calibration systematic error and uncertainty that is less than 1\% in magnitude and 1~deg in phase, as we expect signals with high signal-to-noise ratios in the upcoming observing runs.
    Such standards in calibration will allow for precise and accurate claims about the subtleties of astrophysical source properties derived from gravitational waveforms.
    In order to characterize the uncertainty and systematic error at this level, we strive for our absolute reference network to have precision and accuracy less than 0.1\% in magnitude and 0.1~deg in phase.
    This subsection highlights the work done to-date on achieving such a network.

    \subsubsection{Timing References}

    Timing of the detectors' real-time control system, and the data stored therefrom, including channels used to reconstruct the detector strain, both within and between detectors, is a critical absolute reference.  
    The timing system's custom design is discussed at length in~\cite{bartos2010advanced}.
    Overall and inter-site timing is achieved through the use of Global Positioning System (GPS) receivers, located at the corner and end stations of each interferometer, which provide timing signals to the detector. 
    Although the exact implementation of propagating the GPS timing information to the timestamp on a given bit of data has changed from O1 to O4, the checks and validation on that process have not.  
    Cross checks between multiple GPS clocks and a local atomic clock have been found to provide reliable timing at better than the level of $\pm 1\,\si{\micro\second}$~\cite{sullivan2023timing}.
    This has been and remains entirely sufficient for astrophysics.
    For example, the timing accuracy for a binary neutron star signal in Advanced LIGO with a signal-to-noise ratio of 10 is approximately 0.1 ms \cite{Fairhurst_2011}.
    Further, internal to a detector's timing system, a $\pm 1\,\si{\micro\second}$ timing uncertainty may be recast as a phase delay or advance uncertainty in frequency-response, which only surpasses 1~deg at 2.7~kHz, outside of the most sensitive portion of the detectors' frequency band.

    \subsubsection{Direct Force References}
    \label{sec:dispreferences}
    
    The current generation of ground-based GW detectors employ an auxiliary laser system called photon calibrators (Pcals) as an absolute calibration reference in order to generate the calibrated fiducial displacements.
    Pcals use the radiation pressure from the reflection of power-modulated auxiliary 1047\,nm laser beams to apply a force directly on its given arm cavity optic, displacing it along the arm cavity beam axis~\cite{karki2016advanced}. 
    One Pcal system is installed at each end station on the X and Y arms of the detectors.

    The amplitude of the Pcal-induced displacement is proportional to the amplitude of the modulated Pcal laser power. 
    Thus, the accuracy of the periodic fiducial displacements is directly dependent on the accuracy of the Pcal laser power sensors. 
    A global scheme to characterize the Pcal power sensors has been implemented to improve the accuracy of the absolute calibration in collaboration with national metrology institutes in the US (NIST) and Germany (PTB) and to reduce relative calibration errors between the LIGO and Virgo detectors~\cite{karki2022toward}. 
        
    Improvements in the Pcal measurement techniques and the automation of data analysis processes, along with NIST and PTB's efforts to provide better calibration accuracy, has enabled the generation of fiducial displacements with only 0.3\,\% uncertainty for the current observing run, a significant improvement from 0.42\,\% in O3~\cite{bhattacharjee2020fiducial} and 0.72\% in O1 and O2~\cite{karki2016advanced}. 
    The major source of uncertainty in Pcal-induced displacement is unwanted rotation of the test mass due to the Pcal center of force position offset or an interferometer beam position offset.
    At the LIGO Hanford site, the interferometer beams are deliberately positioned more than 20~mm away from the nominal position at the center of the test mass to minimize the effect from point absorber defects \cite{brooks2021point}. 
    However, at Livingston, the test mass  coatings were improved to remove point absorber defects before the start of O4.
    This allowed the interferometer beam to operate closer to its nominal position, thus reducing the Pcal uncertainty estimate from 0.3\,\% to 0.15\,\%.
    It is expected that similar upgrades to reduce point absorbers on the coatings at the LIGO Hanford observatory would lead to the same level of Pcal uncertainty.
         
    Another method to directly apply a force to the arm cavity mirrors makes use of gravity~\cite{Mio1987, matone2007}. 
    These systems, referred to as the Newtonian calibrator (Ncal) or gravity field calibrators (Gcal), have been developed, tested and integrated at the Virgo~\cite{estevez2018, estevez2021} and KAGRA~\cite{Inoue2018} observatories and have shown promise of delivering displacement at sub-percent level accuracy. 
    Ncals are a system of rotating masses that cause periodic changes in the local gravitational field applying a known gravitational force on the end test masses (ETMs). 
    The relative standard uncertainty on the amplitude of the Ncal-induced displacements at Virgo during O4 is 0.17\,\%, a significant improvement over the previously reported uncertainties with the Ncal prototype~\cite{aubin2024}. 
    LIGO has constructed its own Ncal system, producing time varying forces at twice and three times the Ncal rotation frequency. 
    During O3 it was demonstrated at LIGO Hanford that a calibrated displacement well above the detector sensitivity could be generated using this system to less than 1\% relative uncertainty~\cite{ross2021initial}.
    Using Pcal in conjunction with Ncal systems could help identify systematic errors in these systems and reduce overall displacement calibration uncertainty.

    In addition to directly forcing the arm cavity optics, several indirect methods for inferring displacement have been used in the past with varying levels of success. Discussed further in  \ref{app:indirectdisprefs} for posterity, these methods often rely on a multitude of cascading measurements using tools and detector components which are difficult to characterize well, resulting in inadequate levels of uncertainty.

    \subsubsection{Mass and Dynamical References}
    \label{sec:quaddynamics}

    Critical to any of the direct-force actuator absolute reference methods described above is an understanding of the complex-valued, frequency-dependent, force-to-displacement (and force-to-rotation, torque-to-displacement, and torque-to-rotation) dynamical response of the test mass and its quadruple suspension system. This multi-stage isolation and its force-actuation system is described in detail in~\cite{aston2012update}. 

    The physical parameters of the modeled suspension dynamics include measured, designed, and accepted values and for masses, moments of inertia, lengths, dimensional volumes, lever-arms, and Young's Moduli. 
    Measurements of suspension model parameters have been performed repeatedly for over 15 years, and the dynamics have shown no evidence of time-dependence.
    We also carefully measure the absolute displacement (rotation) per a given force (torque).
    We chose to set this scale with transfer functions that are known not to be particularly sensitive to any but the most well-measured parameters.
    Studies like that from~\cite{karki2016advanced} and \cite{ross2021initial} have shown that in the 10 to 50 Hz region, treating the multi-stage system as though the test masses were an entirely free inertial mass has sufficiently low systematic bias even at the 0.1\% level.
    The $\sim$40 kg masses are measured on scales (in-air) that are cross-checked with calibrated mass standards to the accuracy of 10 g.
    In frequency regions where the dynamical response is not so simple, both the Pcal and Ncal systems may be used to directly measure what complex (and complex-valued)  dynamical response may be present across the entire detection band, validating the model further.

    \subsection{Differential Arm Length (DARM) Loop Model}
    \label{sec:darmmodel}

    Constructing a model for the DARM loop is crucial to producing an accurate and precise calibration.
    The DARM loop is broadly divided into two components: (1) the interferometric response of the dual-recycled Michelson interferometer with \FP cavities to differential displacement of the interferometer arms measured by photodetectors, known as the sensing function $C$, and (2) the feedback signals controlling the position of the test masses as part of the multi-stage quadruple pendula, known as the actuation function $A$.
    Together, these components form the DARM loop, or detector, response that we model as $R^{(\textrm{model})}$ in order to estimate $\Delta L_\textrm{free}$.
    The model of the detector response and estimation of the systematic error and uncertainty on measured strain have evolved in complexity since the detection of GW150914.
    A full description of the current models for the actuation and sensing functions can be found in \ref{app:darmloop}.

  The mathematics to describe the signals measured by the interferometer DARM loop have been discussed in detail in~\cite{abbott2017calibration,cahillane2017calibration,sun2020characterization}.
  Here, we provide a brief summary of the basics of the physical model. 
    When the detector is operating in its lowest-noise configuration, the DARM loop digital error signal, $d_{\textrm{err}}$, is linearly proportional to differential displacement of the arms, $\Delta L_\textrm{free} \equiv \Delta L_{x} - \Delta L_{y}$. 
    The true response function, $R$, may be treated as a simple, single-input-single-output (SISO) control loop: the error signal is a suppressed version of the external disturbance, which can be expressed in the frequency domain as
    \begin{eqnarray}
    \tilde d_{\textrm{err}}(f) & = & \frac{C(f)}{1 + G(f)}~\tilde{\Delta L_\textrm{free}}(f)  
    \label{eq:loopsupp}  
    \end{eqnarray}
    where $G$ is the feedback loop's open loop gain transfer function, $1 / (1+G)$ is the loop suppression transfer function, and $C$ is the sensing function.
    The differential arm displacement in units of physical length may be recast as strain on the detector, $h_{\textrm{det}} = \Delta L_\textrm{free} / L$, where $L$ is the average length of the arms\footnote{Note, this is \emph{not} the astrophysical strain, $h_{\textrm{astro}}$, incident upon the detector; the relationship between astrophysical strain and detector strain involves a set of source- and frequency-dependent functions~\cite{schutz1987antenna, tinto1987antenna, rakhmanov2008high}. }.
    Using \eref{eq:loopsupp}, we can solve for the more physically interesting quantity $h_{\textrm{det}}$, once again written in the frequency domain,
    \begin{equation}
    \tilde h_{\textrm{det}}(f) = \frac{1}{L}~R(f)~\tilde d_{\textrm{err}}(f) \label{eq:hoft_fromderr}
    \end{equation}
    with the detector response function defined as
    \begin{equation}
    R(f) \equiv \frac{1 + G(f)}{C(f)} \ . \label{eq:Rdef}
    \end{equation}
    Due to where the physically interesting quantity enters the DARM loop, the complexity of the DARM loop actuation system, and the inability to measure displacement without the entire detector control system running, the actuation function, $A$, is segregated from the sensing function, $C$, and the open loop gain is defined as the product,
    \begin{equation}
    G(f) \equiv A(f)~D(f)~C(f) \label{eq:Gdef}
    \end{equation}
    where $D$ is a collection of well-known digital filters that creates a digital control signal $d_{\textrm{ctrl}} \equiv D~d_{\textrm{err}}$.
    The Pcal is used while the detector is operating in its low-noise configuration to continuously measure and validate the modeled quantities for $C^{(\textrm{model})}$ and $A^{(\textrm{model})}$, which are used to create a model of the response function $R^{\textrm{(model)}}$ via \eref{eq:Rdef} and \eref{eq:Gdef}.

    The software used to construct all of the above components of the DARM loop model has evolved significantly since the detection of GW150914. 
    The software implementation calculates frequency response values of DARM loop components forming $A^{\textrm{(model)}}$ and $C^{\textrm{(model)}}$.
    In O1-O2, the model was constructed in Matlab, as this was the native software for the most complex portion of the loop -- the quadruple pendulum dynamics.
    Calibration products that rely on the model, however, were constructed in Python, utilizing Python packages including filter generation and Gaussian Process Regression for estimating unknown systematic error (see \sref{sec:strainproduction} and \sref{sec:syserrbackgnd} for further details).
    
    By O3, the modeling software was entirely Python-based in a package called \texttt{pyDARM} \cite{pydarm}.
    Currently during O4, \texttt{pyDARM} is a modular, flexible Python package, publicly available with continuous integration unit tests.
    Specifically, \texttt{pyDARM} now allows modeling of more complicated loop topologies, interfacing with control room computers to facilitate measurements and analysis of results, rapid deployment of filters when an update to the calibration is required, and careful record-keeping of model changes and measurements during observing runs.

    \subsection{Producing the Calibrated Strain Data}
    \label{sec:strainproduction}    
    
    With a detector response model in hand, the detector strain $h_{\rm det}$ can be constructed in the time domain or in the frequency domain either by convolving the digital error signal $d_{\rm err}$ with a digital filter representing $R$ (time domain) or by a direct multiplication of $\tilde d_{\rm err}(f)$ with $R(f)$ (frequency domain).  The result should be divided by the unperturbed arm length to achieve units of strain (see \eref{eq:hoft_fromderr}).
    Alternatively, the detector strain can be constructed from a combination of the digital error and digital control signals, either in the frequency or time domain.  The LIGO calibration has historically been done by constructing the detector strain in the time domain from a combination of the digital error and digital control signals,
    \begin{equation}
        h_{\rm det}^{(\textrm{model})}(t) = \frac{1}{L}\left(\frac{1}{C^{(\textrm{model})}} * d_{\rm err}(t) + A^{(\textrm{model})} * d_{\rm ctrl}(t) \right) \ ,
        \label{eq:hoft}
    \end{equation}
    where here $*$ references a convolution operation and $C$ and $A$ are expressed as time-domain digital filters.  As introduced above, the superscript ``model" references the fact $C$ and $A$ are based on constructed models, rather than the exact (and typically unknown) detector sensing and actuation functions.  This choice -- to use $d_{\rm err}$ and $d_{\rm ctrl}$ (two proportional versions of the same signal, with the proportionality exactly known) -- was originally made in the initial LIGO (iLIGO) era \cite{siemens2004making} in order to account for slow time dependence of $A$ and $C$ in a straight-forward manner and to remain insensitive to changes in the digital system between those two points (namely, changes in $D$, which are often required to improve the noise of the detector).
    Details on how the detector strain is constructed using digital filters can be found in \cite{viets2018reconstructing, wade2023improving} and is discussed in more detail in \sref{subsec:pipeline}.

    \subsection{Systematic Error and Uncertainty}
    \label{sec:syserrbackgnd}

    The constructed detector strain, $h_{\rm det}^{\textrm{(model)}}$, must be accompanied by a well-quantified systematic error estimate and uncertainty on the systematic error estimate.
    As the detector response is changing on the order of minutes, it requires estimating systematic error in $h_{\textrm{det}}^{\textrm{(model)}}$ on the same timescale.
    The systematic error at any given time is modeled in the frequency domain.
    We estimate the quantity
    \begin{equation}
        \eta_{R}(f) = \frac{R(f)}{R^{\textrm{(model)}}(f)}~,
        \label{eq:etaR}
    \end{equation}
    which is equivalent to $h_{\textrm{det}}(f) / h_{\textrm{det}}^{\textrm{(model)}}(f)$ (from \eref{eq:hoft_fromderr}).
    Note that $\eta_R(f)$ is a multiplicative factor to scale $h_\textrm{det}^{\textrm{(model)}}(f)$ to $h_\textrm{det}(f)$, where $h_\textrm{det}(f)$ is the exact detector strain that calibration is striving to reconstruct.
    A preliminary version of this equation
    appears as equation (13) in~\cite{abbott2017calibration}, and is later refined in equation (5) of~\cite{cahillane2017calibration}, and solidified in equations (9) -- (11) of~\cite{sun2020characterization}.

    At the time of GW150914, the systematic error was modeled in a frequentist framework in the fashion of the iLIGO error model \cite{abadie2010calibration}.
    The response function was analytically broken down into partial derivatives of its components to form a ``propagation of errors'' assuming independence of all measurements or statements that informed the production of $A^{\textrm{(model)}}$ and $C^{\textrm{(model)}}$; see equations (7) and (8) of~\cite{cahillane2017calibration}. 
    It became apparent, however, that the detector response was more complex than anticipated and our measurements of seemingly independent parameters showed correlation.
    Inspired by astrophysical parameter estimation techniques which employ numerically evaluated Bayesian statistics to characterize uncertainty, we began to create an analysis framework for modeling systematic error that appropriately handled many cross-correlated parameters with physics-agnostic algorithms necessary for the accurate representation of the error.
    This framework, described in~\cite{sun2020characterization}, has been used throughout O1-O3 to model systematic error on the reconstructed detector strain.  
    During O4, this framework has evolved using newly available measurements, described in \sref{sec:currentwork}. 
    
    Here we summarize the framework used in O1-O3 as it still remains within the O4 framework. 
    Using \eref{eq:hoft}, we observe that $R$ is composed of $C$ and $A$, and conclude that $\eta_R$ is also composed of $\eta_C$ and $\eta_A$.
    Section 3.2 of \cite{sun2020characterization} discusses how the Pcals may be used to directly measure $A$ and $C$ for any given time, and thus build a measurement of $\eta_{C}$ and $\eta_{A}$ at the limit of Pcal uncertainty.
    We find that these terms are dominated by the systematic error between the models and measurements of $C$ and $A$.
    Section 4.3 of~\cite{sun2020characterization} describes the use of Gaussian Process Regression (GPR) methods~\cite{JMLR:v12:pedregosa11a} to fit the measured systematic error in $\eta_{C}$ and $\eta_{A}$.
    Crucially, the GPR provides a posterior distribution of the fits of $\eta_{C}$ and $\eta_{A}$ that we sample numerically as part of the estimate of $\eta_{R}$.
    This results in a posterior distribution of $\eta_R$ at each selected short interval of time\footnote{This is typically done during a 2-minute interval at an hourly cadence while the detector is in a mode suitable for observation, but the process can be run at specific times, if desired.}.
    Although we implicitly assume that successive measurements of $\eta_{C}$ or $\eta_{A}$ are drawn from the same parent distribution, in reality the detectors evolve in time, and our models for $C$ and $A$ may not fully reflect the evolution.
    Nevertheless, kernel hyperparameter values, which encode information about the shape and behavior of the covariance function used in the GPR, are chosen such that the measurement results are contained within the 1-$\sigma$ percentile of the distribution of curves instantiated from the GPR kernel.

    Figures~10 and 11 of~\cite{sun2020characterization} illustrate examples of systematic errors of the response function model.
    In \ref{app:calibsyserrorlimits}, we review some of the reasons for these kind of \emph{measured} errors.
    The GPR-informed method of \emph{modeling} the measured systematic error over-estimates $\eta_{C}$ and $\eta_{A}$ in some frequency regions, and does not capture the complexity of the error in other frequency regions.
    These two issues continue to dominate the model of $\eta_R$.
    Said differently for clarity -- the value (frequency-dependent magnitude and phase) of the systematic error and uncertainty in the calibrated data, $\eta_{R}$, has not been limited by the uncertainty of the absolute reference network that is used to characterize it; instead by the limitations of choices in modeling its underlying features.

    In~\cite{abbott2017calibration}, we published  the model of systematic error in the reconstructed strain data at the time of GW150914.  
    By the end of O1, the number of transient GW detections were still low enough to again publish the specific systematic error estimates for the detector strain at the time of each detection, see Appendix A of~\cite{cahillane2017calibration}.
    During O2, although astrophysical parameter searches still used event-by-event systematic error estimates, it became necessary to quantify the systematic error estimate over several entire epochs of observation time~\cite{cahillane2017calibration}.
    The procedure remained in place for O3~\cite{sun2020characterization, sun2021characterization}, with seven epochs for the Hanford detector and four epochs for the Livingston detector. Epochs are defined by major detector changes throughout the run.
    The measurement sets that define $\eta_{R}$ within those epochs -- the measured systematic error in $\eta_{C}$ and $\eta_{A}$ -- may have entirely different response after an epoch boundary. 
    As such, the measurement collection is ``reset'' at every boundary and we do not expect $\eta_{R}$ to improve linearly with time.
 
    During O4, advances in handling DARM loop configuration changes and the processes incorporating those changes into the response model relaxes the firm epoch boundary constraints from O1-O3, aside from exceptional DARM loop changes.
    In addition, continuous measurement of $\eta_{R}$ at specific frequencies (see \sref{sec:cal-monitoring}) permits a measurement approach that improves the accuracy of the modeled estimate of calibration systematic error.
    These improvements will be presented and discussed in a forthcoming publication.

    Delivering the model of systematic error has historically taken several weeks to months, due to the complexity of combining measurements, accounting for detector changes, and addressing larger known systematics.
    During O1-O3, the strain data reconstructed in low-latency was not accompanied by a systematic error estimate (see figure~\ref{fig:evolution}).
    Instead, the systematic error estimate was determined from a revised version of the strain data generated offline, with latency on the order of months.
    During O4, we have addressed many of the challenges so that systematic error estimates are now provided alongside the low-latency calibrated strain data.

\section{Anticipated Challenges for LIGO Calibration in O5}
\label{sec:cal_challenges}
The decade following the discovery of GW150914 has brought significant advancements in LIGO calibration efforts and has moved us from a place of providing final calibration data products (strain data and accompanying systematic error estimates) on timescales of months towards timescales of hours.  
Although, the process of reviewing and certifying calibration products still takes weeks to months.
Moving forward, the LIGO calibration effort will need to continue to evolve in tandem with ongoing improvements to the LIGO detectors.
Detector upgrades will impact the DARM loop that is at the heart of calibration, and requirements on the accuracy and latency of the calibrated data will continue to tighten as louder and more gravitational-wave sources become observable.

The planned detector upgrades for O5 that will significantly impact calibration include the implementation of a Balanced Homodyne Detection (BHD) readout scheme \cite{fritschel2014balanced}, replacement of the test masses and an increase to the circulating arm power from 400\,kW to 750\,kW \cite{fritschel2025instrument}.
The BHD readout scheme will introduce an adjustable homodyne angle, the audio-band demodulation quadrature of
the GW readout, which will need to be included in the sensing function model~\cite{hall2019systematic}.
The higher circulating arm power, if implemented without additional thermal actuation, will lead to larger thermal transients in the sensing response of the detectors each time the detectors transition from a not-operational (cold) state to the full-power (hot) operating state. 
Thermal actuation systems, described more below, or a more sophisticated sensing model will need to be developed to mitigate these large thermal transients.
The replacement of test masses will require an update to the models of suspension dynamics, which directly impacts the actuation function.
The test mass upgrade also has the potential for resetting the number and location of point absorbers, which influence both the sensing response and, indirectly, the amount of cross-coupling in the actuation dynamics.

The planned upgrade to circulating arm power of 750 kW is likely to be particularly impactful on the calibration in O5.
As the arm cavity mirrors heat up during the thermalization period, they experience surface deformation due to absorption from the main interferometer beam. 
This deformation will affect the frequency-dependent sensing response, $C(f)$. 
However, the current sensing function model is not sophisticated enough to account for thermalization effects in real time. 
The detector strain reconstructed during each thermalization period will then contain errors due to the mismatch between the sensing function model used for reconstruction and the true interferometer sensing function during the thermalization period. 
The interferometer is said to be thermalized once the the circulating arm power reaches roughly 95\% of its nominal operating value and the thermal transient is observed to resign. 
In O4, that period was roughly 1.5-2\,hours.
Figs.~\ref{fig:monplots_time} and~\ref{fig:monplots_violin} in \sref{sec:cal-monitoring} show the effect of thermalization on the systematic error of the calibrated data during O4.

To mitigate a frequency-dependent systematic error in the sensing function, $\eta_{C}$, introduced by the evolving sensing response during the thermalization periods, we need to develop noise-free methods to track the sensing function and employ a time- and power-dependent model of the sensing function that can be used to reconstruct  the detector strain in real time as the detector response evolves. 
In early O4, low-frequency, sinusoidal injections from the Pcals were used to provide a known reference for the detector strain at a frequency where thermal transient effects likely dominate ($6-15$\,Hz) \cite{sun2021characterization}. 
However, these tests were discontinued because it was found that the injections might have been inadvertently causing noise at higher frequencies.
One possibility is to explore improvements to how these low-frequency Pcal injections can be implemented that avoids unwanted noise at other frequencies.

Another approach to addressing the negative impacts of thermalization on the sensing function is to reduce the physical impact of thermalization on the detector response.  
There are ongoing efforts to deploy additional heating actuators
that will effectively heat the test mass surfaces to mitigate the
effects of thermalization during power up. These are the The FROnt Surface Type Irradiator (FROSTI) \cite{FROSTI} 
and Central Heater for Transient Attenuation (CHETA) \cite{capote2025advanced}.
In particular, the CHETA actuator promises to significantly mitigate the majority of  calibration-affecting thermalization issues by using a CO$_2$ laser source pointed at the cavity-facing surfaces of the input and end test masses to maintain the mirrors in a ``thermalized'' state via absorption at times when the circulating arm power is $\sim$0\,W. 
Once successfully deployed, the CHETA project will significantly shrink the gap between the ``cold'' and ``hot'' interferometer states, rendering thermalization effects negligible.

As discussed in \sref{sec:syserrbackgnd}, estimates for LIGO calibration systematic error rely on measurements of the DARM loop taken at specific moments in time.
For gravitational wave signals that are broadband and transient in nature, e.g., a binary neutron star coalescence, the broadband estimate of calibration systematic error can be mapped in a straightforward way to this waveform, and it is safe to assume the calibration systematic error is static during the measurement of the gravitational wave signal. 
On the other hand, for signals that are narrow-band and long-duration in nature, e.g., rapidly-rotating neutron stars emitting a weak, continuous, quasi-monochromatic gravitational-wave signal, the calibration systematic error might not be well-accounted by the broadband, static estimate. 
Continued work is required in order to properly estimate calibration systematic error for narrow-band signals and to integrate those time-varying estimates into posterior distributions of source parameter estimates.

Overall, as the LIGO detectors become more sensitive and as the signals these detectors are capturing become more varied, the requirements on calibrated data products will continue to evolve.  To meet this challenge, we need to establish more flexible, real-time methodologies for both generating calibrated strain data and estimating the systematic error on the calibrated data.

\section{Advancements in LIGO Calibration Accuracy, Latency, and Monitoring}
\label{sec:currentwork}
Throughout O4, we have implemented improvements to existing infrastructure and developed a suite of new tools for improving the monitoring, latency, and reliability of calibrated data in LIGO.  These innovations mark advancements towards meeting calibration requirements expected in O5 and beyond. 

Due to the increased rate of GW detections \cite{detection-rate}, the improvements in pipeline reliability discussed below, the development of a real-time monitoring system on the low-latency calibrated data, and the development of a low-latency estimate of calibration error and uncertainty, the decision was made at the beginning of O4 to use the low-latency strain data as the final calibrated strain data product whenever possible.  Prior to O4, final published results of analyses were based on a second, high-latency calibration produced from archived data, months after initial data acquisition.  This was done primarily to ensure accurate knowledge of the calibration models and a well-informed estimate of the systematic error and uncertainty in those models.  Since the beginning of O4, a high-latency calibration is produced only occasionally for time periods with known significant errors.  For O4a and O4b, which spanned approximately 20 months of observing time, a total of about 2 months of data required high-latency recalibration.

\subsection{Low-latency Calibration Pipeline Improvements}
\label{subsec:pipeline}

The low-latency calibration pipeline used from O1--O4 is divided into two stages, the first of which is accomplished using the front-end computers used to control the detector.  The front-end calibration uses infinite impulse response (IIR) filtering to compute an estimate of the strain in each detector.  The front-end calibration pipeline's exclusive use of IIR filters and limitation to timestamping data only in real time lead to deficiencies in the pipeline's ability to accurately represent the detector response.  The second stage of calibration is accomplished in the \texttt{gstlal} calibration pipeline~\cite{viets2018reconstructing,cannon2021gstlal}, a \texttt{GStreamer}-based pipeline that wraps LIGO Algorithm Library (\texttt{LAL})~\cite{lal} software with \texttt{GStreamer}'s audio streaming capability.  The \texttt{gstlal} calibration pipeline was used from O1--O4 to produce calibrated data in real time using partially calibrated data from the front end as input, as well as in high latency using archived raw data as input.  The \texttt{gstlal} calibration pipeline uses time-domain finite impulse response (FIR) filters and is therefore able to apply calibration models with arbitrary frequency dependence and produce strain data that is independent of the pipeline's start time, a necessity for exactly reproducing strain data.  The \texttt{gstlal} calibration pipeline also computes a bitwise state vector whose primary purpose is to indicate times when the strain data is suitable for analysis.  

\subsubsection{Improvements to Calibration Pipeline Latency}
\label{subsubsec:pipelineLatency}

The contribution of the calibration pipeline to the latency of LIGO strain data has been monitored since the beginning of O2, when the calibration latency was $\sim$13 seconds.  Due primarily to two improvements, the calibration latency was reduced to $\sim$3 seconds by the beginning of O3 and has remained at this values throughout O4. 
To put this latency in perspective, the first preliminary alert for public gravitational-wave events throughout O4 has an expected median latency of $\sim30$ seconds \cite{LowLatencyAlerts}.
One source of latency in the calibration pipeline that was eliminated was caused by anti-aliasing and anti-imaging filters associated with downsampling and upsampling processes in the measurement, calculation, and application of the time-varying parameters of the DARM model.  Each of the two such resampling processes added $\sim$6 seconds of latency.  This $\sim$12-second latency was eliminated by including an option in the resampling algorithm to simply shift the timestamps of the resampled data.  With no latency contribution from this branch of the pipeline, the FIR filters, which occupy a parallel branch of the pipeline, became the dominant source of pipeline latency.

The remaining latency has been minimized by significant improvements to the quality of the FIR filters used in the \texttt{gstlal} calibration pipeline to apply the models $1/C^{\rm (model)}$ and $A^{\rm (model)}$.
The FIR filters are computed by taking the inverse Fourier transform of a model that is represented in the frequency domain.  The resulting time-domain filter is then windowed to optimize its frequency response.
Due to seismic noise at low frequencies, the filters include a very aggressive high-pass roll-off below 10 Hz.  Achieving adequate high-pass filtering without incurring a large latency in the strain data is one of the primary challenges in FIR filter design.  During O2,
the filters were windowed using a Tukey window with a cosine fraction of 0.9.  Using this window, the high-pass filtering stipulation required long filters that incurred 4 seconds of latency.  By the beginning of O3, a more fine-tuned window design had reduced the filter latency to 2 seconds while also improving the quality of the high-pass filter.  In O4, a Slepian window is used, resulting in filters with better attenuation of low-frequency noise and tunable frequency resolution.  Small delays associated with data transfer into and out of the pipeline and with all the computations done in the pipeline add $\sim$1 second, leading to the current total latency of $\sim$3 seconds.

\subsubsection{Improvements to Pipeline Reliability and Robustness}
\label{subsubsec:pipelineReliability}

Since O1 and O2, the calibration pipeline has also become significantly more reliable and robust.  One of the main issues that has been thoroughly addressed was caused by frequent dropouts of raw data at the input to the \texttt{gstlal} calibration pipeline.  These dropouts can occur in any subset of the $\sim$100 input channels and can vary in length from seconds to hours, but rarely exceed a few seconds during observation.  A method was introduced at the beginning of O2 to fill in the data dropouts with zeros.  The calibration state vector indicates times when these dropouts corrupt the strain data.  By the beginning of O4, more flexible options were included in order to minimize loss of data quality in the output.  Dropouts can be filled with: (1) zeros - the most common option, indicating data that should not be used in strain production if possible, or that the strain data computed is not suitable for analysis, (2) user-defined nonzero values - useful for instances in which the pipeline needs to handle data dropouts in a different manner than actual zeros in a channel, or (3) the last good value - useful for channels containing constants that rarely change.
Additionally, during real-time calibration, the pipeline uses a timer to determine when a dropout is occurring.  Prior to this improvement, made during O3, the pipeline could only identify a dropout based on the timestamp of the first input data after the dropout ended.  Long dropouts therefore caused the pipeline output to incur latency, causing the delayed strain data to be dropped in the data aggregation processes that follow calibration.  This is no longer an issue, as the pipeline continuously fills in missing data during long dropouts.

\subsubsection{Development of an Exact Solution for Time Dependence in the Calibration Model}
\label{subsubsec:exactTDCFs}

The models $C^{\rm (model)}$ and $A^{\rm (model)}$ include time-dependent parameters that can vary on timescales ranging from $\sim$minutes to $\sim$weeks.  
See \eref{eq:sensingfunction} and \eref{eq:actuationfunction} for reference on time-dependent parameters in the sensing and actuation function models.  
The time-dependent parameters in the sensing function include $\kappa_C$, $f_{\rm cc}$, $f_{\rm s}$, $Q$ and $\tau_C$. The time-dependent parameters in the actuation function include $\kappa_i$ and $\tau_i$ where $i = (U,P,T)$ is the upper suspension stage $(U)$, the penultimate suspension stage $(P)$ and the test mass suspension stage $(T)$.
Methods to track and compensate for the time dependence in these models during the production of strain data were initially developed during O1 \cite{tuyenbayev2016improving}.  These methods were incorporated into the \texttt{gstlal} calibration pipeline during O1 to produce a high-latency calibration that included compensation for the gain factors $\kappa_i$.  Beginning at the start of O2, compensation for the $\kappa_i$ was included in the low-latency calibration as well, and a high-latency calibration additionally included compensation for the coupled cavity pole of the sensing function $f_{\rm cc}$, achieved using adaptive filtering methods described in~\cite{wade2023improving}.  Compensation for the $\kappa_i$ and $f_{\rm cc}$ has been included in the low-latency calibration since the beginning of O3.  Compensating for time-dependence has reduced our median systematic error from $\sim$10\% and $\sim$$10$ deg in magnitude and phase, respectively, for O1's low-latency strain data product to $\sim$3\% and $\sim$$5$ deg in magnitude and phase, respectively, for O3's low-latency strain data product~\cite{cahillane2017calibration, sun2020characterization}.  Methods have also been developed to track and compensate for the parameters $f_{\rm s}$ and $Q$ of the sensing function and $\tau_i$ of the actuation function, but work is still ongoing to study the impact of compensating for these parameters during strain reconstruction.

The methods currently used to compute time-dependent parameters in $C^{\rm (model)}$ and $A^{\rm (model)}$ rely on simplifying approximations, resulting in $\sim$1\% errors in the estimates of time-dependent parameters.  The out-of-phase addition of the actuation and sensing components of $h_{\rm det}^{\rm (model)}$ shown in \eref{eq:hoft} inflates these errors in the strain data at some frequencies.  An ``exact'' solution, described in chapter 5 of \cite{viets2019optimizing} has been developed in the \texttt{gstlal} calibration pipeline but was not yet implemented in the production of strain data during O4 as review and validation are still ongoing.  
This method will likely be implemented in O5, eliminating a systematic error that is currently uncompensated in the production of strain data.
These systematic errors are largely accounted for—though not explicitly modeled—when estimating the overall calibration error and uncertainty budget in O4. 
The implementation of the exact solution may also produce more consistent improvements to calibration accuracy with the application of the full time-dependent model.

\subsubsection{Subtraction of Spectral Lines}
\label{subsubsec:lineSubtraction}

Searches for GWs are negatively impacted by narrow-band noise sources in the strain data \cite{davis2021}.  
Some narrow-band noise sources in the form of persistent, sinusoidal signals, known as calibration lines, are intentionally added via the Pcal lasers and the actuation system, in part to measure the time-dependent parameters of the calibration models during observation.  
A method for subtracting spectral lines, including the calibration lines, in low latency was implemented in the \texttt{gstlal} calibration pipeline starting at the beginning of O3.  The method relies on the use of witness channels, which are correlated to the signal to be subtracted but insensitive to GWs.  This is intended to prevent the unintentional subtraction of GWs.  The method is carried out by first demodulating strain $h(t)$ and the witness channel $w_j(t)$ at the frequency of the $i^{\rm th}$ line $f_i$ to find complex values $\tilde{h}(f_i)$ and $\tilde{w}_j(f_i)$ which encode the amplitude and phase of $h$ and $w_j$ at that frequency.  A transfer function value is then computed at $f_i$ using a running median as
\begin{equation}
    T_j(f_i)={\rm med}\left(\frac{\tilde{h}(f_i)}{\tilde{w}_j(f_i)}\right)
\end{equation}
If there are multiple witness channels $w_j$ at frequency $f_i$, transfer function values are found by solving
\begin{equation}
    W^{jk}T_j=V^k
\end{equation}
with the matrix $W^{jk}={\rm med}\left(\frac{\tilde{w}^j} {\tilde{w}_k}\right)$ and the vector $V^k={\rm med}\left(\frac{\tilde{h}}{\tilde{w}_k}\right)$.  Summation over the index $j$ is implied.  The cleaned signal is then computed as
\begin{equation}
    \tilde{h}_{\rm clean}(f) = \tilde{h}(f)-\sum_i\sum_j T_j(f_i) \tilde{w}_j(f_i) \ .
\end{equation}
In order to prevent the addition of latency, the final subtraction is carried out in the time domain by reconstructing the subtraction signal as a sum of sinusoids.

\begin{figure}
    \centering
    \includegraphics[width=0.9\linewidth]{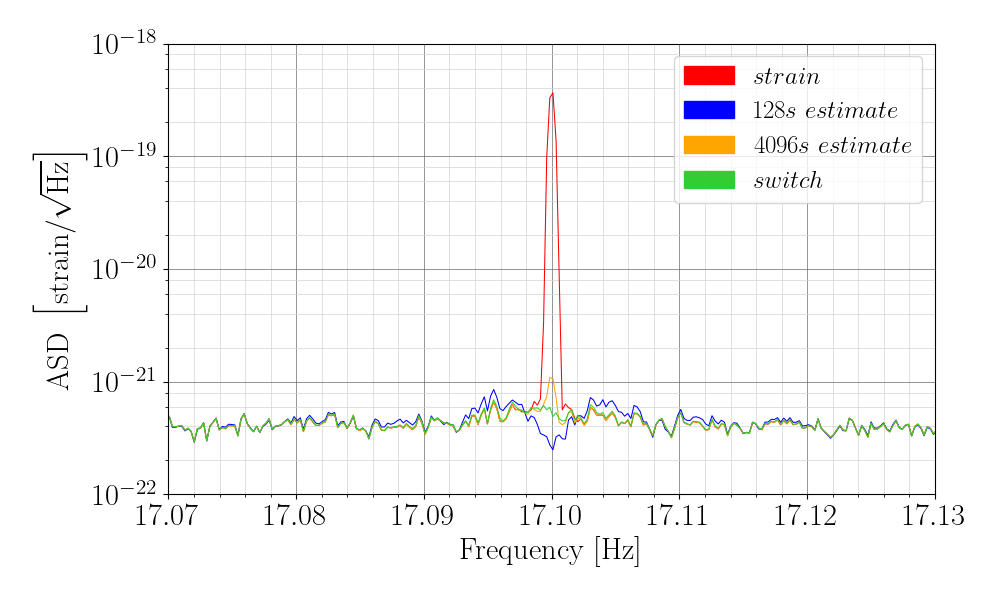}
    \caption{An example of line subtraction from data in the LIGO Hanford detector.  This figure shows the amplitude spectral density (ASD) zoomed in on the 17.1 Hz Pcal line.  The data with no line subtraction applied is shown in red.  The data with line subtraction performed using a static 128 second running median method is shown in blue.  The data with line subtraction performed using a static 4096 second running median method is shown in orange.  The adaptive switching method that dynamically changes the length of the running median based on whether a secular change is detected in the $T_j$ before taking the median is shown in green.  The adaptive switching method yields the best results and is the method implemtened during O4.}
    \label{fig:linesubtraction}
\end{figure}

During the review of calibrated strain data in O3 and O4, several issues and shortfalls of the low-latency line subtraction performed by the calibration pipeline were identified.  These issues were addressed and fixed before the final release of the O3 and O4 strain data.  One problem that was addressed during O4 is a narrow-band systematic error, often referred to as ``over-subtraction,'' introduced near the line frequencies.  This error is caused by the use of a finite amount of data to inform the transfer function values $T_j$, leading to an imperfect ability to distinguish between signal that is present in the witness $w_j$ and signal that is not.  The bandwidth of this error is inversely proportional to the length of the running median, which had been set to 128 seconds during O3.  Increasing the length of the running median mitigates the error, but it also hinders the ability to subtract spectral lines when the $T_j$ are evolving on shorter timescales.  During O4, a method was implemented that allows the subtraction algorithm to dynamically change the length of the running median based on whether significant secular change is detected in the $T_j$ before taking the median.  The length of the running median is now adjusted between a minimum of 128 seconds and a maximum of 4096 seconds.  Figure~\ref{fig:linesubtraction} demonstrates the subtraction of the 17.1 Hz Pcal line using the 128 second running median, the 4096 second running median, and the adaptive switching method.

Numerous types of occasional line subtraction failures have been caused by data dropouts and by insufficient communication of detector state information to the subtraction algorithm.  Many of these have been addressed by more sophisticated handling of data dropouts and by more thoroughly utilizing all available state information.  The O4 implementation of line subtraction uses multiple metrics to determine when the input data is suitable for measurement of the $T_j$ and when the calibration lines are turned on or off.  With the addition of the Pcal lines used for calibration monitoring in O4 (discussed below), the line subtraction performed in the low-latency calibration procedure has become increasingly essential to the production of high-quality strain data for downstream analyses.  The improvements to the quality and robustness of the line subtraction algorithm have been essential to this achievement.

\subsection{Real time Calibration Monitoring}
\label{sec:cal-monitoring}

The current observing run, O4, has brought the first deployment of a real-time monitoring system for calibrated LIGO data, \texttt{CalMonitor}. 
This development is one step along the road towards achieving low-latency, high-accuracy calibrated data that is robust to a changing interferometer.
\texttt{CalMonitor} makes use of calibration lines injected by the Pcals.
While Pcal calibration lines have been used throughout all of Advanced LIGO for various purposes, O4 marks the first time these lines have been analyzed in low-latency with results published to a live web application for real-time monitoring and review.
Additionally, we are now injecting additional Pcal lines that are independent of the strain computation and serve only as monitors of calibration accuracy at specific frequencies.
The frequencies corresponding to each of the calibration lines used for monitoring are known as monitoring line frequencies $f_{mon}$.
At the monitoring line frequencies, the strain of the detector is dominated by the Pcal injection, $\tilde h_{\rm det}(f_{mon}) \approx \tilde h_{\rm pcal}(f_{mon})$.  
The signal-to-noise ratio of these injected signals is such that a 256 second integration window yields an uncertainty in the injected signal coherence with the calibrated strain data of less than 2\%.  
A real-time measure of the calibration systematic error $\eta_R$ at the monitoring line frequencies can then be constructed from the transfer function of the Pcal injected signal $h_{\rm pcal}$ in units of strain and the reconstructed detector strain data $h^{(\rm model)}_{\rm det}$ at each monitoring line frequency,
\begin{equation}
    \label{eq:calsyserror}
    \eta_R(f_{\rm mon}) = \frac{\tilde h_{\rm pcal}(f_{\rm mon)}}{\tilde h^{(\rm model)}_{\rm det}(f_{\rm mon})} \ .
\end{equation}
The generally complex-valued $\eta_R$ at the monitoring line frequencies would be unity in the case of perfect detector calibration.

\begin{figure}
    \centering
    \includegraphics[width=0.9\linewidth]{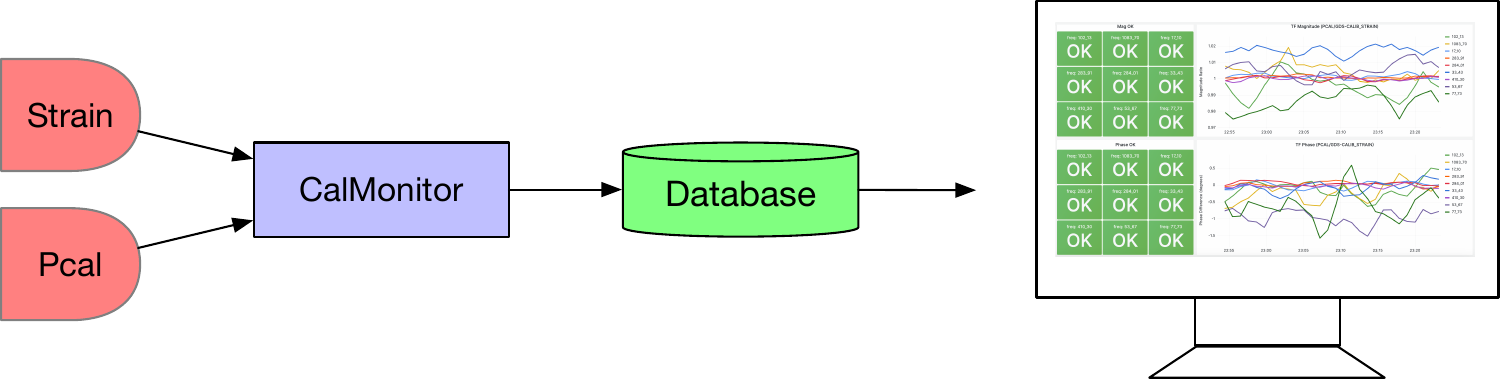}
    \caption{Simplified workflow for the data products associated with \texttt{CalMonitor}.  The red containers indicate timeseries data streams that are required as inputs to \texttt{CalMonitor}.  The data products produced by \texttt{CalMonitor} are broadcast using the Kafka event streaming platform and then aggregated and stored in an InfluxDB backend using the \texttt{ligo-scald} software package.  The details of these steps are omitted from this summary diagram, but the end result of calibration metric aggregation in a database is represented by the green cylinder.  The Grafana web application is used to query InfluxDB and display calibration metrics in a real-time, user-friendly interface.}
    \label{fig:calmonitor}
\end{figure}

\texttt{CalMonitor} uses the streaming software library \texttt{gstlal} to perform real-time computations of the calibration systematic error at the monitoring line frequencies \cite{cannon2021gstlal}.
In addition to calibration systematic error, we also construct and store metrics related to the state of the instrument, coherence of each transfer function measurement, values for the time-varying DARM  model parameters \cite{tuyenbayev2016improving, viets2019optimizing,}, and metrics related to the Pcal system, discussed more below.
All of these constructed metrics are broadcast using the Kafka event streaming platform.  
The LVK software package \texttt{ligo-scald} is used to store these metrics from Kafka in InfluxDB \cite{ligo-scald}. 
We make use of the Grafana web application to visualize the real-time calibration metrics.
\Fref{fig:calmonitor} shows the data flow for the real time calibration monitoring system.
In addition to visualizing calibration metrics using the Grafana web application, the transfer function measurements stored in InfluxDB are also pulled into our hourly calibration systematic error estimates, as will be discussed more in a forthcoming publication.

\begin{figure}[htbp]
    \centering

    \begin{subfigure}[b]{0.48\textwidth}
        \centering
        \includegraphics[width=\textwidth]{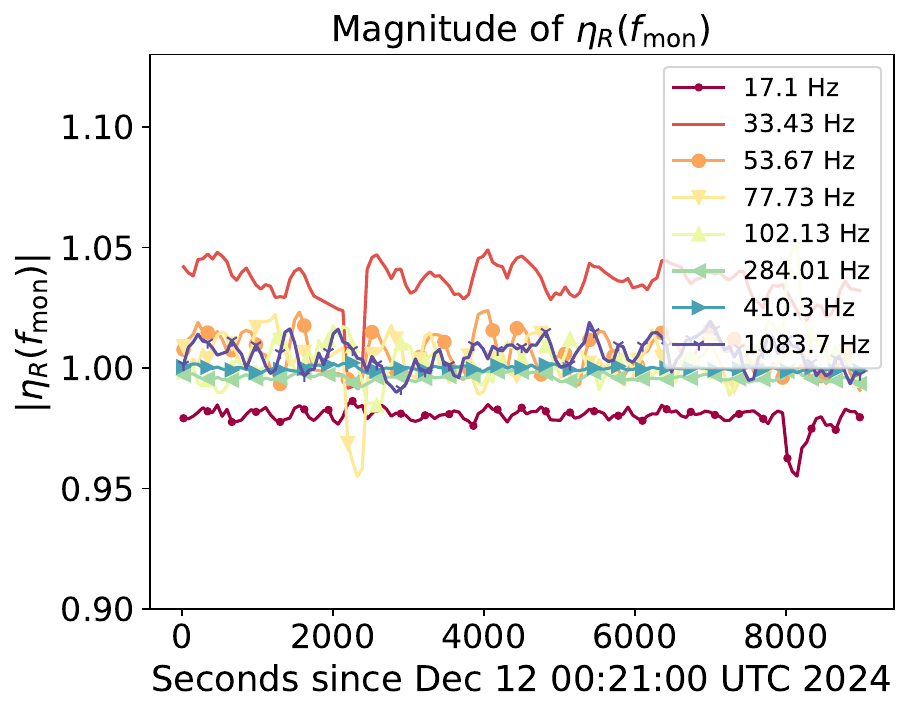}
    \end{subfigure}
    \hfill
    \begin{subfigure}[b]{0.48\textwidth}
        \centering
        \includegraphics[width=\textwidth]{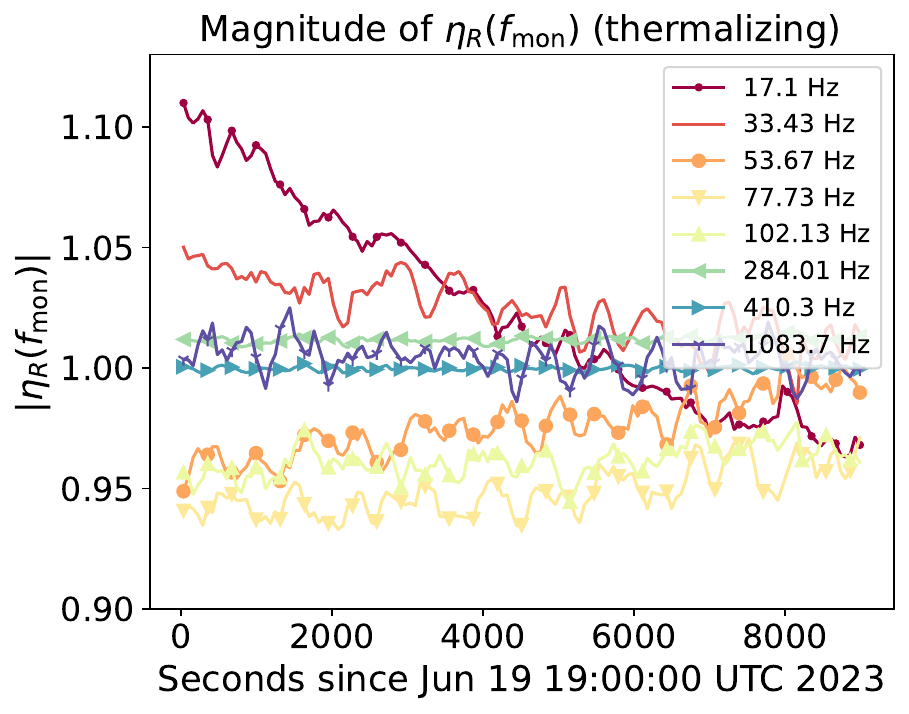}
    \end{subfigure}

    \vskip\baselineskip

    \begin{subfigure}[b]{0.48\textwidth}
        \centering
        \includegraphics[width=\textwidth]{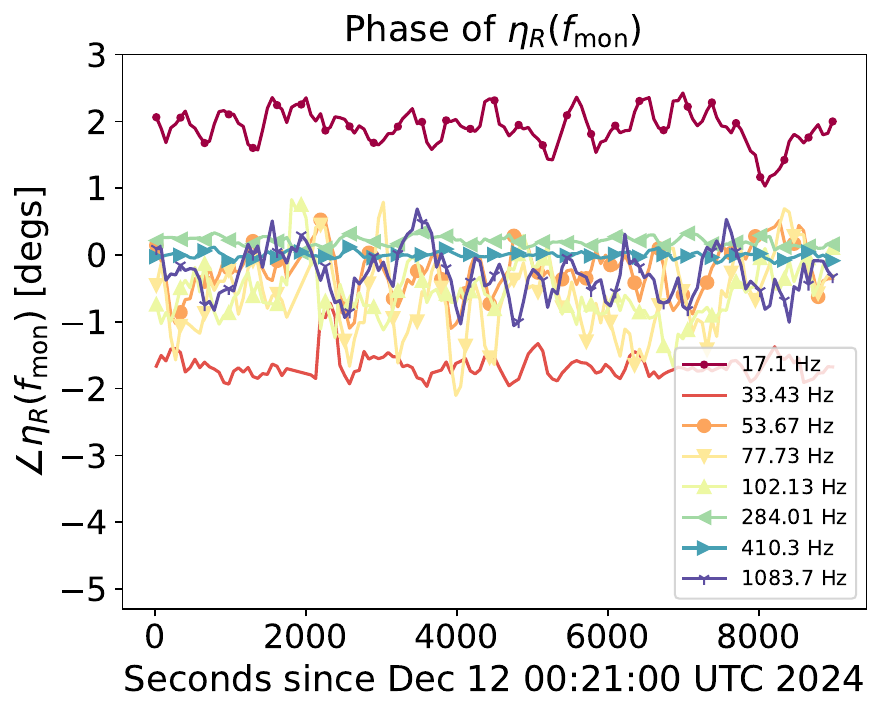}
    \end{subfigure}
    \hfill
    \begin{subfigure}[b]{0.48\textwidth}
        \centering
        \includegraphics[width=\textwidth]{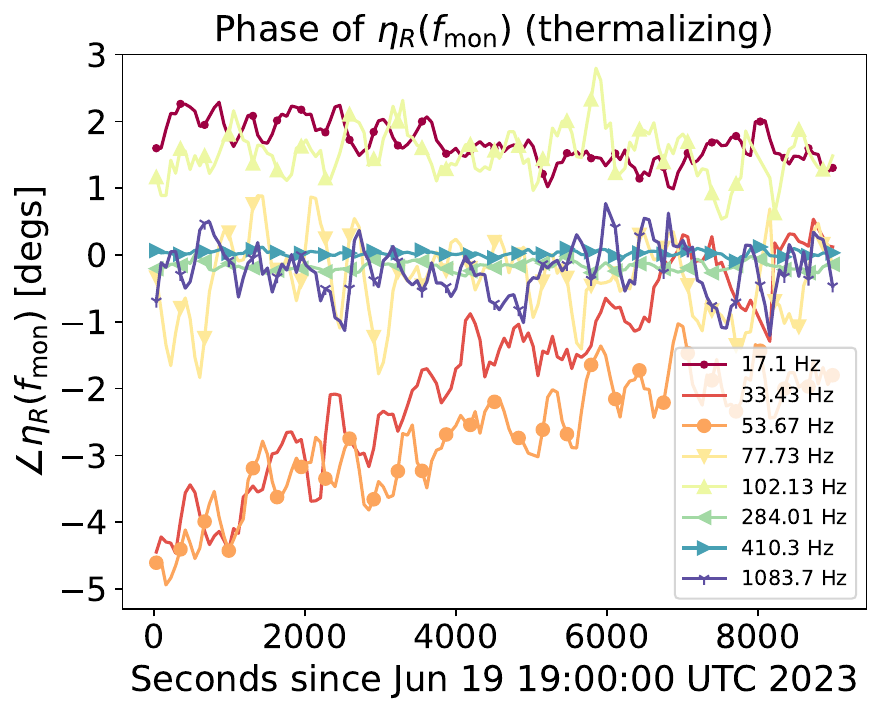}
    \end{subfigure}

    \caption{Time series of the magnitude and phase of calibration systematic error as measured by the monitoring lines (see \eref{eq:calsyserror}) in the LIGO Hanford detector.  The top plots show the magnitude of the calibration systematic error and the bottom plots show the phase.  The plots on the left are for a time period when the detector was fully thermalized and no thermalization effects were measured in the detector sensing function.  The plots on the right are for a time when the detector was actively thermalizing and these effects are measured in the low-frequency monitoring lines as a relatively large systematic error in the calibration.}
    \label{fig:monplots_time}
\end{figure}

\begin{figure}[htbp]
    \centering

    \begin{subfigure}[b]{0.9\textwidth}
        \centering
        \includegraphics[width=\textwidth]{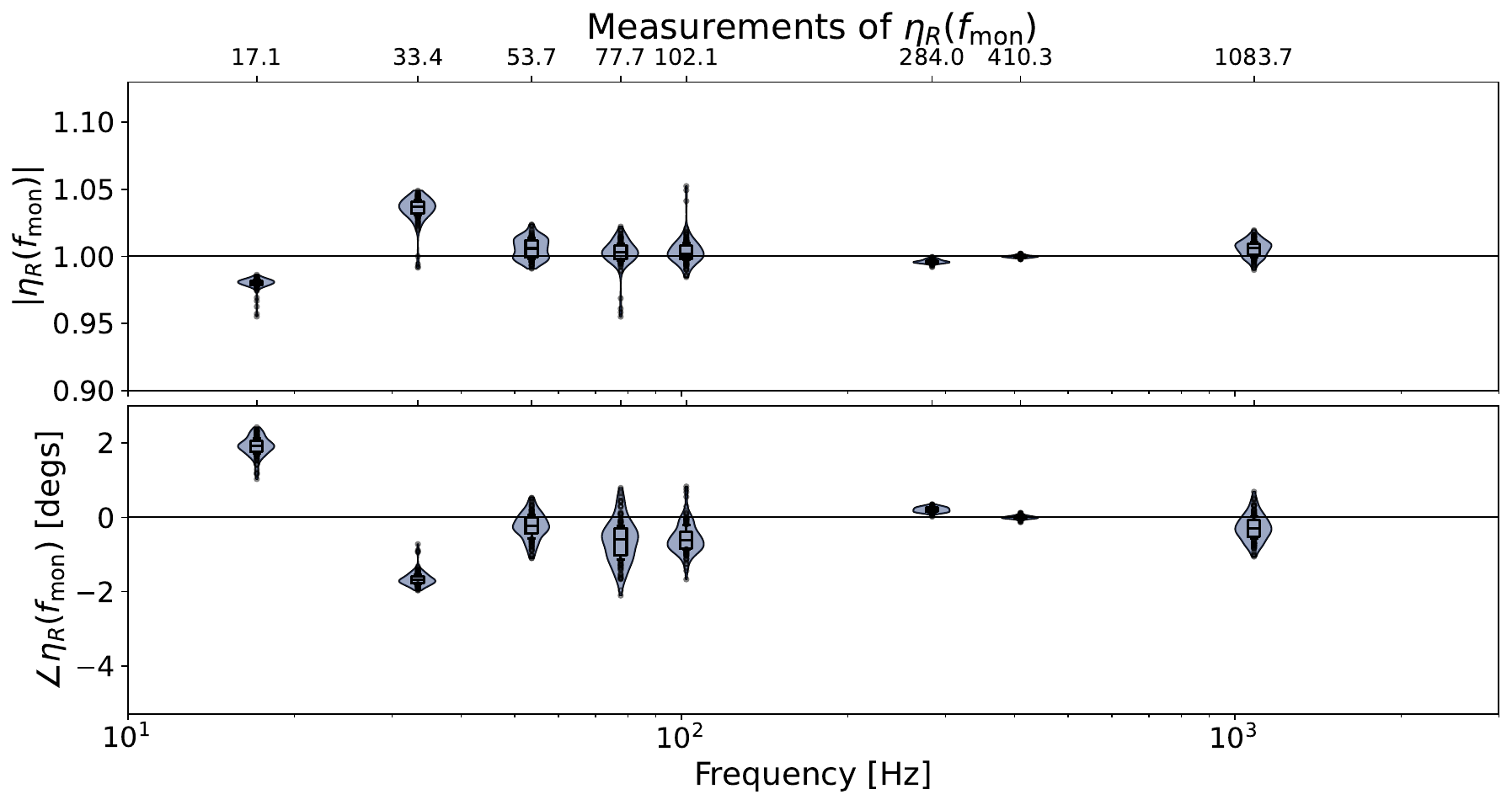}
    \end{subfigure}
       
    \vskip\baselineskip
        
    \begin{subfigure}[b]{0.9\textwidth}
        \centering
        \includegraphics[width=\textwidth]{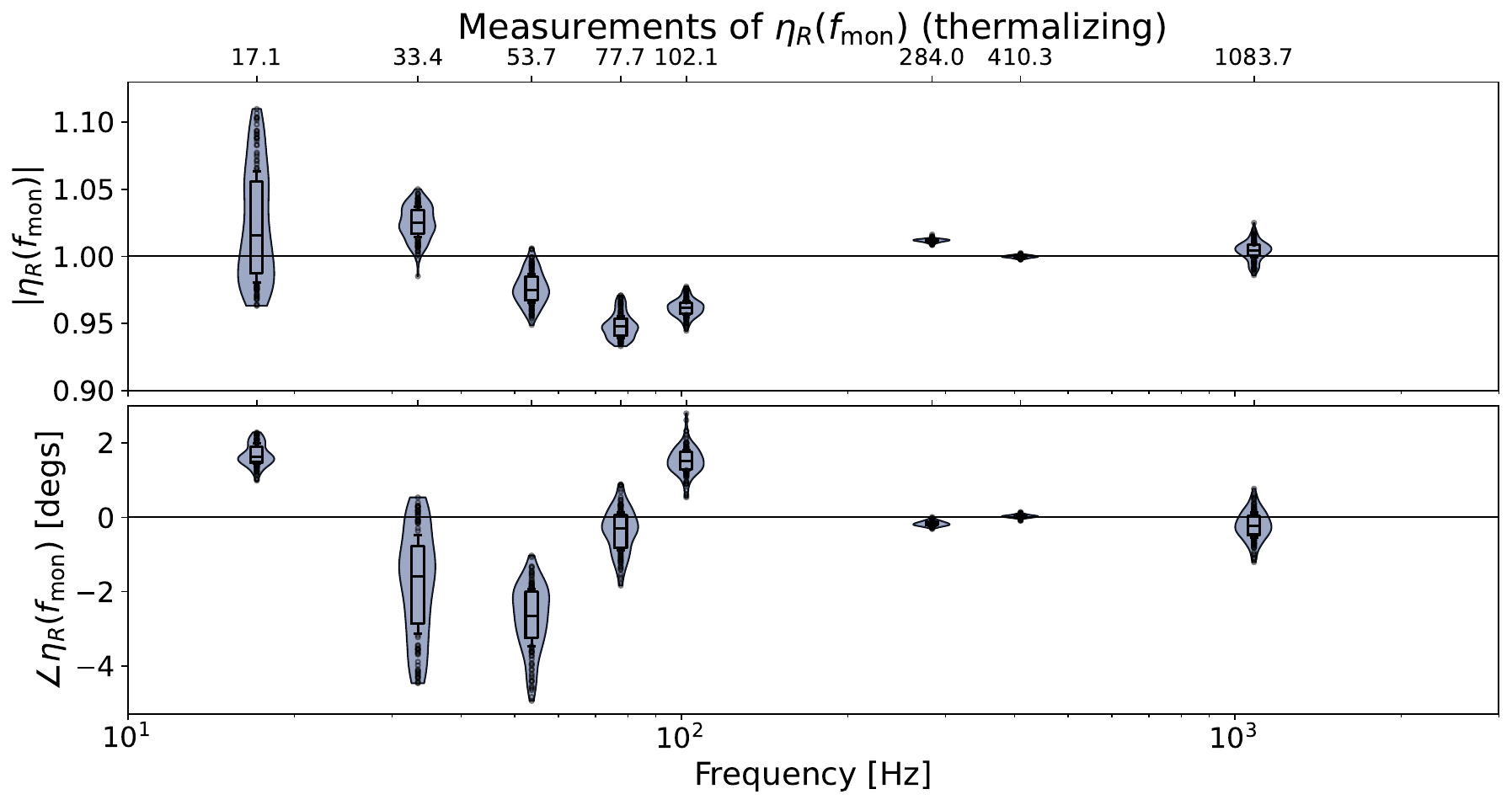}
    \end{subfigure}

    \caption{Violin bode plots of the magnitude and phase of calibration systematic error as measured by the monitoring lines (see \eref{eq:calsyserror}) at the LIGO Hanford detector over the same time period as figure~\ref{fig:monplots_time}.  The top plots show the systematic error in the calibration when the detector was fully thermalized and no thermalization effects were measured in the detector sensing function.  The bottom plots show the systematic error for a time when the detector was actively thermalizing.  The larger spread in systematic error at the lower frequency monitoring lines is a result of the larger systematic error in the sensing function model during thermalization.  The frequencies of the monitoring lines are shown along the top of each bode plot.  The box on each violin plot extends from the first to third quartile.  The black line shows the median value.  The whiskers stretch from the box to the most distant data point that falls within 1.5 times the interquartile range of the box. Any points beyond the whiskers are considered outliers.}
    \label{fig:monplots_violin}
\end{figure}

Figures~\ref{fig:monplots_time} and~\ref{fig:monplots_violin} are representative of the types of plots displayed in real time on the Grafana web application.  These plots display point estimates at specific frequencies for the systematic error of the calibrated data over the selected time period, as computed through \eref{eq:calsyserror}, a violin plot representing the magnitude and phase of the systematic error of the calibration as a function of frequency over the selected time period, and time series of the time-varying DARM model parameters.

The implementation of the real-time calibration monitoring system has allowed for quick identification of problems with LIGO calibrated data and sanity checks on our computed calibration systematic error estimates.  
For example, in the early months of O4, the LIGO Hanford detector was operating with a higher laser power.  The thermalization effects resulting from the higher laser power led to systematic errors in our modeled sensing function, as discussed in \sref{sec:cal_challenges}.  The real-time calibration monitoring system clearly showed the trend of larger systematic errors during times of thermalization, as shown in figures~\ref{fig:monplots_time} and~\ref{fig:monplots_violin}.  The effects are particularly apparent at the lowest frequency monitoring lines, where the thermalization effects have the largest impact on the detector sensing function.

\begin{figure}[htbp]
    \centering
    \includegraphics[width=0.8\textwidth]{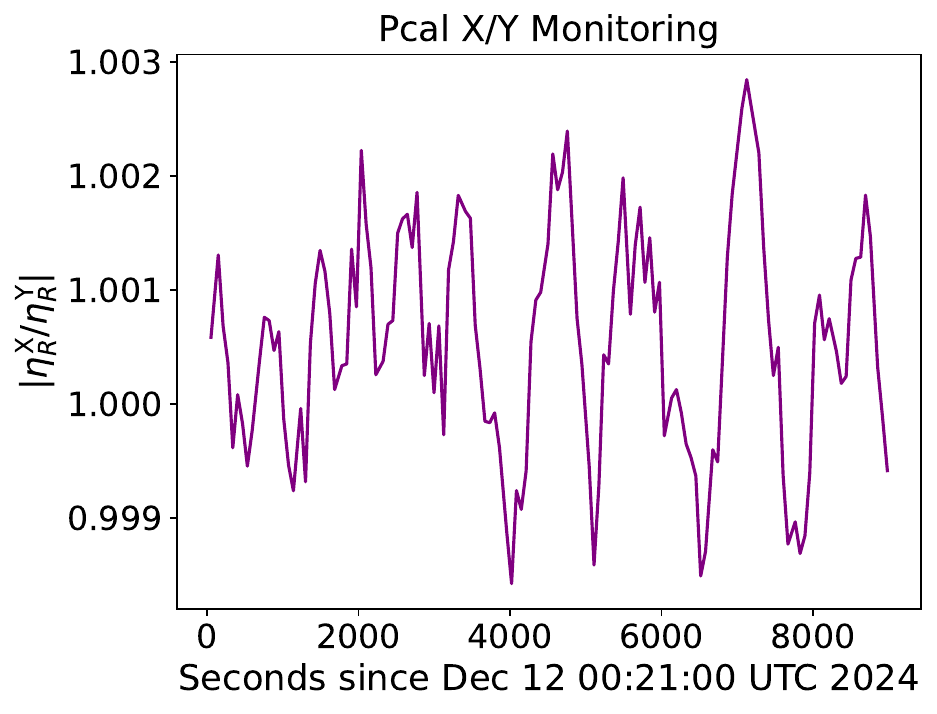}
    \caption{Time series of the magnitude of the characteristic ratio $\eta^{\rm X}_R / \eta^{\rm Y}_R$ (see \eref{eq:pcalxy}) at the LIGO Hanford detector used to identify any relative errors between the two Pcal calibrations at each end station. 
 This ratio should be one for identical Pcal calibrations. The deviation from unity is due to uncertainties in the contributing factors in Pcal calibrations, not common to the two end stations.  The data in this plot was collected using a configuration where $\eta^{\rm X}_R / \eta^{\rm Y}_R$ was computed using three 50\% overlapping 240 second FFTs. The mean of the variations in the ratio is $\sim 1.0005$, indicating that there is no significant relative Pcal errors. }
    \label{fig:pcalxy}
\end{figure}

The real-time calibration monitoring system has also been employed to study the Pcal absolute references.  
Using the characteristic feature of the interferometer that it cannot distinguish whether the X-end or the Y-end test mass was displaced, the two Pcal signals at each end station can be combined to provide a more accurate absolute calibration. 
This formalism was developed in O3~\cite{bhattacharjee2020fiducial}. 
During O4, two frequencies $f_{\rm X}$ and $f_{\rm Y}$, excited $0.1$ Hz apart, were continuously monitored at both the Hanford and Livingston sites. 
Figure \ref{fig:pcalxy} shows a time series of the quantity $\eta^{\rm X}_R / \eta^{\rm Y}_R$ for the LIGO Hanford detector, where
\begin{equation}
    \eta^{\rm X,Y}_R(f_{\rm X,Y}) = \frac{\tilde h^{\rm X,Y}_{\rm pcal}(f_{\rm X,Y})}{\tilde h^{(\rm model)}_{\rm det}(f_{\rm X,Y})} \ 
\end{equation}
and $\tilde h^{\rm X,Y}_{\rm pcal}$ refers to the injected signal by the X-,Y-end Pcal in units of strain.  Or, more directly,
\begin{equation}
    \label{eq:pcalxy}
    \frac{\eta^{\rm X}_R}{\eta^{\rm Y}_R} = \frac{\tilde h^{\rm X}_{\rm pcal}(f_{\rm X})}{\tilde h^{\rm Y}_{\rm pcal}(f_{\rm Y})} \ ,
\end{equation}
where we assume $\tilde h^{(\rm model)}_{\rm det}(f_{\rm X}) \approx \tilde h^{(\rm model)}_{\rm det}(f_{\rm Y})$ as $f_{\rm X}$ and $f_{\rm Y}$ are closely spaced.
Continuous monitoring of $\eta^{\rm X}_R / \eta^{\rm Y}_R$ enabled the identification and correction of any relative error between the two Pcal calibrations at the end stations, ensuring an accurate absolute calibration in real time. 

The real-time calibration monitoring system became a critical tool for the LIGO calibration team during the O4 observing run.
The infrastructure for \texttt{CalMonitor} was designed to be flexible enough to allow for expanded calibration metrics and monitoring in future observing runs.
Similar systems will be required as we move into the next-generation of GW detectors as well.

\section{Calibration for the Next Generation of Gravitational-wave Detectors}
\label{sec:nextgen}
The next-generation GW observatories, such as Cosmic Explorer~\cite{evans2021} and Einstein Telescope~\cite{Maggiore2020}, will extend GW observations to unprecedented distances, probing fundamental physics and astrophysical processes with unparalleled sensitivity. To fully realize their potential, robust and highly accurate calibration methods must be developed to ensure that raw interferometric data are precisely translated into astrophysical strain measurements \cite{PhysRevD.105.082002, Capote:2024mqe}.

Real-time calibration will be a crucial component of next-generation observatories. 
Future instruments will require real-time, high-accuracy calibration to enable immediate astrophysical analyses and better facilitate rapid responses for early-warning alerts, which are alerts that are sent out before the merger phase of the binary coalescence, and multimessenger observations in a high signal-to-noise ratio regime \cite{PhysRevD.97.123014}. 
Achieving this requires advancements in detector calibration technology and methodology, including improved hardware, more sophisticated real-time monitoring, and efficient computational algorithms to interpret measurements and rapidly adjust the calibration.
Alternative calibration schemes, such as astrophysical calibration, also provide promising methods for calibrating the next generation detectors \cite{Essick_2019, Schutz:2020hyz}.

Achieving the necessary calibration accuracy presents significant technical challenges. Advanced LIGO and Virgo have achieved systematic uncertainties on the order of a few percent, with significant effort in improving the DARM model in high latency, but next-generation detectors will require calibration errors to be reduced potentially well below the 1\% level in magnitude and to within a fraction of a degree in phase~\cite{evans2021}. This level of accuracy is critical for precise measurements of the neutron star equation of state, stringent tests of general relativity, and improved constraints on cosmological parameters such as the Hubble constant.

To meet these rigorous requirements, several key advancements are under development. Pcals must achieve significantly improved precision, potentially reaching uncertainties below 0.1\%. Alternative calibration references, such as Newtonian calibrators, may offer new pathways to achieving ultra-precise displacement measurements~\cite{aubin2024}. Additional methodologies, including modulated laser frequency techniques and auxiliary interferometric reference systems, may be used to complement Pcals and provide independent verification of calibration accuracy~\cite{evans2021}.

Beyond hardware improvements, real-time computational framework will play a central role in next-generation calibration. Future observatories will require fully integrated, real-time calibration solutions that continuously account for detector response variations. This will demand advances in real-time data processing algorithms, machine learning techniques for rapid anomaly detection, and improved feedback control systems to ensure stable calibration over extended observing runs.

The development of these next-generation calibration methodologies builds upon the extensive progress made during the Advanced LIGO-Virgo-KAGRA observing runs. 
By refining systematic error modeling, improving real-time response characterization, and expanding the range of independent calibration techniques, the field is steadily advancing toward the goal of high-precision, real-time calibration. 
These innovations will be instrumental in unlocking the full scientific potential of Cosmic Explorer, Einstein Telescope, and other future GW observatories, ensuring that they provide the most accurate and reliable data for astrophysical and fundamental physics discoveries.

\section{Conclusion}
\label{sec:conclusion}

Over the past decade, LIGO calibration has moved from producing calibrated strain data and the associated systematic error estimate on these data on a time scale of several months to a timescale of seconds for the data and hours for the systematic error estimate.
During the time of GW150914, we were able to achieve calibration errors less than $10\%$ in magnitude and $10$ deg in phase in the detection band.  However, this required many months of work to produce recalibrated strain data from a well-understood DARM model.
In O4, we are producing calibrated strain data with the same level of accuracy as was done for GW150914 but now with significantly lower latency. 
The low-latency strain data are available within seconds of data collection and preliminary uncertainty estimates are available on an hourly cadence.
However, a thorough review of all data products is still a time-consuming process.
The shift towards low-latency calibrated data that have a well-quantized systematic error estimate has eliminated the need for computationally expensive and time-consuming reruns of astrophysical analyses on recalibrated strain data that are delivered months after data collection.

The improved accuracy and latency of the calibrated strain data has been possible due to advances in the low-latency calibration pipeline, including improvements to the digital filters used in the calibration process and the subtraction of narrow-band noise in the calibrated strain data. 
Additionally, the development of a real-time monitoring system for the calibration has allowed for quick identification of problems in the calibration and provided real-time measurements of the calibration systematic error that inform the systematic error estimates.  
The real-time calibration monitoring system has also been deployed to measure any unexpected, systematic differences between the Pcals at the X- and Y-end stations.
The real-time calibration monitoring system will be expanded to include new calibration metrics, such as metrics related to line subtraction and thermalization effects.

As we move into O5 and look towards calibration with future gravitational-wave detectors, the need for even higher accuracy and lower latency calibrated strain data will become evident.
We must continue to develop systems that can make real-time measurements of the GW detectors to enable fast and accurate calibrated strain data without interfering with astrophysical observations.
Additionally, any future system for GW detector calibration must build in the ability to deploy calibration error estimates alongside the strain data in real time.
Monitoring will be a critical component of real-time calibration systems in order to ensure the rapid identification of issues within the calibration processes.
Calibrated strain data is at the heart of GW science derived from this data, and vigilance to ensure the accurate and speedy delivery of calibrated data will further enable the accurate and speedy delivery of the science that flows from this data.

\section{Acknowledgments}
Calibration of the LIGO detectors has always required an eye towards detail and a deep understanding of the GW detectors.  
Many folks have contributed meaningfully to LIGO calibration from initial LIGO to the present.  
This list of individuals includes, but is not limited to, Vladimir Bossilkov, Craig Cahillane, Virginia d'Emilio, Jenne Driggers, Becca Ewing, Jane Glanzer, Gaby Gonzalez, Evan Hall, Chad Hanna, Hsiang-Yu Huang, Yuki Inoue, Kiwamu Izumi, Dana Jones, Shivaraj Kandhasamy, Sudarshan Karki, Keita Kawabe, Mike Landry, Greg Mendell, Brian O'Reilly, Avani Patel, Ethan Payne, Jameson Rollins, Richard Savage Jr., Xavier Siemens, Darkhan Tuyenbayev, Gabriele Vajente, and Alan Weinstein.
This material is based upon work supported by NSF's LIGO Laboratory which is a major facility fully funded by the National Science Foundation (NSF).
LIGO was constructed by the California Institute of Technology and Massachusetts Institute of Technology with funding from the United States NSF, and operates under cooperative agreement PHY–2309200. Advanced LIGO was built under award PHY–0823459.
The authors gratefully acknowledge the support of the United States NSF for the construction and operation of the LIGO Laboratory and Advanced LIGO as well as the Science
and Technology Facilities Council (STFC) of the United Kingdom, the Max-Planck-Society
(MPS), and the State of Niedersachsen/Germany for support of the construction of Advanced LIGO and construction and operation of the GEO600 detector. 
Additional support for Advanced LIGO was provided by the Australian Research Council (ARC).
The authors are grateful for computational resources provided by the LIGO Laboratory and supported by NSF Grants PHY-0757058 and PHY-0823459.
MW, LW, and EM are supported by NSF grants PHY-2308796.  
MW, DB, MC and AV are supported by the NSF grants OAC-2103662.
EG is supported by the Canada Foundation for Innovation and the National Sciences and Engineering Research Council of Canada.
LS is supported by the Australian Research Council (ARC) Centre of Excellence for Gravitational Wave Discovery (OzGrav), Project Number CE230100016, and the ARC Discovery Early Career Researcher Award, Project Number DE240100206. 

\appendix

\section{Indirect Displacement References}
\label{app:indirectdisprefs}

    Prior to the advanced LIGO detectors, the technology for establishing an absolute reference that could cause a calibrated differential arm displacement that was both independent from the detector and performing at noise levels that did not impact the sensitivity of the detector did not exist. 
    As such the field began by using the detector itself and its properties to infer the differential displacement indirectly or through a cascading series of measurements, while such ``direct'' technologies were under development.

    Over the past decades, GW interferometers have implemented a variety of techniques to generate calibrated fiducial displacements, which are used to characterize and correct for the effects of the feedback control loop~\cite{goetz2010accurate}. 
    During initial LIGO and even the early period of Advanced Virgo, the free-swinging Michelson method~\cite{abadie2010calibration} was used to provide the absolute calibration of the detectors. 
    This technique uses the interferometer laser wavelength as a length reference and the measurement of Michelson interference fringes to calibrate the end test mass (ETM) actuation. 
    Another method that did not require any localized forces to be impinged on the test mass to calibrate the ETM actuation is a frequency modulation method~\cite{goetz2010calibration}. 
    These methods, however, require the interferometer to be in a ``single-arm" configuration rather than its nominal configuration. 
    Propagating the calibration at its nominal configuration at a sub-percent uncertainty level proved to be a challenging task using these methods.

\section{The DARM Loop}
\label{app:darmloop}
The DARM loop, as introduced in \sref{sec:background}, is the feedback control loop that stabilizes the differential arm (DARM) degree of freedom in the interferometer.  
The response of the DARM loop, commonly called ``the detector response", must be  characterized in order to achieve accurately calibrated differential arm length data from the interferometer.  
Ever since the km-scale LIGO detectors have begun operation, the parameters of the complex-valued, frequency-dependent detector response are observed to vary slowly over time and change the detector response for three primary reasons:
(1) the alignment of the core optics (suspended from sophisticated multi-stage seismic isolation system in the Advanced LIGO era) drifts on the time-scale of seconds even while under servo control;
(2) the core optics absorb a fraction of the circulating laser power in the \FP cavities (of order watts) and thus distort under thermal load, changing the radii of curvature, and thus the detector response at various minute-long timescales; and 
(3) new for the advanced LIGO era, the alignment of the primary arm cavity optics are controlled using electrostatic actuators which are influenced by residual stray ions within the vacuum enclosure and accumulates over day-long timescales

The two main components of the DARM loop, as discussed in \sref{sec:background}, are the sensing function and the actuation function.  
The models we use for these two components are described in more detail in the following subsections.

\subsection{Sensing Function Model}
    The modeled sensing function has evolved in complexity since GW150914 as the interferometer input laser power and circulating arm power has increased~\cite{abbott2017calibration, cahillane2017calibration, viets2018reconstructing, sun2020characterization}.
    Our current model of the sensing function is a complex-valued, frequency-dependent function given at time $t$ by
    \begin{equation}
    C^{\textrm{(model)}}(f;t) = \kappa_{C}~\left(\frac{H_{C}}{1 + \textrm{i}f/f_{cc}}\right) \left(\frac{f^2}{f^2 + f_{s}^{2} - \textrm{i} f f_{s} / Q}\right) C_{R}e^{-2\pi\textrm{i}f\tau_{C}} \label{eq:sensingfunction}
    \end{equation}
    where $\kappa_{C}$ is a slow time-varying, real-valued scale factor;
    $H_{C}$ is the overall scale of the sensing function; 
    $f_{cc}$ is the so-called cavity pole frequency of the combined \FP and signal recycling cavities~\cite{izumi2016advanced,hall2019systematic}, slowly varying in time; 
    the second parenthetical term is a second-order phenomenological estimate of the measured response below 30 Hz\footnote{This term encapsulates possible arm-to-signal-recycling cavity detuning coupled with non-ideal readout phase and is parameterized by $f_s$ and $Q$.}, where $f_s$ and $Q$ slowly vary in time; 
    $C_{R}$ is any residual frequency response remaining in the true response of the signal processing electronics that cannot be compensated in the near real-time system; 
    and $\tau_{C}$ accounts for the overall time delay.
    The time delay has three contributing components: a phase correction for approximating the full high-frequency response of the \FP cavity~\cite{rakhmanov2002dynamic, rakhmanov2004characterization} as a single pole~\cite{izumi2016toward}, the light-travel time it takes to convey displacement of the end test masses down the arms, and any residual discrete delays incurred from sharing signals across the near-real-time control network of computers. 
    The parameters in this model that are measured as a function of time are $\kappa_{C}$, $f_{cc}$, $f_s$, $Q$, and $\tau_C$.

\subsection{Actuation Function Model}
The modeled actuation function has evolved in complexity together with the physical system since GW150914~\cite{abbott2017calibration, cahillane2017calibration, viets2018reconstructing, sun2020characterization}.
Our current model of the actuation function is a complex-valued, frequency-dependent function given at time $t$ by
\begin{equation}
    A^{\textrm{(model)}}(f;t) = \sum_{i} \kappa_{i}~F_{i}~H_{i}~A_{i}(f)~e^{-2\pi \textrm{i} f \tau_{i}} \label{eq:actuationfunction}
\end{equation}
where $i=(U,P,T)$ is the upper suspension stage ($U$), the penultimate suspension stage ($P$), and the test mass suspension stage ($T$); 
$\kappa_i$ is a slow time-varying, real-valued scale factor; 
$F_i$ is a set of digital IIR filters;
$H_i$ is the overall scale of the actuation of suspension stage $i$;
$A_i(f)$ is a complex-valued, frequency-dependent function of the the combination of (1) the force on suspension stage $i$ to displacement of the test mass transfer function response and (2) any residual complex-valued, frequency-dependent response of uncompensated electronics;
and $\tau_i$ is the overall time delay of stage $i$.

\Eref{eq:actuationfunction} becomes even more complex if several different quadruple suspension optics are used for actuation of the DARM loop.
In principle, one could choose to construct a sophisticated actuation scheme that utilizes quadruple suspension \FP cavity optics of both input and end mirrors;
different stages of different quadruple suspensions could be used or not;
the relative strength on a given quadruple suspension stage could be increased or decreased digitally using the digital control system;
and (adding to the complexity) there are four actuators per suspension stage, where different actuators could be manipulated separately using the digital system.
At the time of GW150914, the DARM loop actuation was intentionally kept simple, but since then, the actuation scheme has slowly grown in complexity so that the actuation function model has also grown in complexity.
A generalized, arbitrary actuation function model is beyond the scope of this article.

In \cite{abbott2017calibration}, equation (4) used the symbol $\mathcal{K}_{i}$ instead of $H_{i}$ to represent the overall scale. 
    Slow temporal variations $\kappa_{i}$ were left out of the definition of $A^{\textrm{(model)}}$ as they were not yet a part of the filters that produced $h_{\rm det}^{\textrm{(model)}}$ at the time. 
    Their measured values were instead incorporated into the systematic error model, $\eta_{R}^{\textrm{(model)}}$. 
    We imposed the expectation that $\kappa_{U}$ and $\kappa_{P}$ did not vary in time and could be lumped together as a common multiplicative factor, $\kappa_{PU}$.
    Finally, it was assumed that each stage had the same computational delay, $\tau_{i} = \tau_{A}$.

    In the offline reproduction of $h_{\textrm{det}}^\textrm{(model)}$ -- for both O1 and O2 and the publication of \cite{cahillane2017calibration} --  the time dependence of the actuation model was parameterized with $\kappa_{\rm T}$, $\kappa_{\rm PU}$, and $\tau_A$.
    By O3, independent values for $\kappa_{i}$ and $\tau_{i}$ were measured and successfully deployed in $A^{\textrm{(model)}}$ and became part of the standard set of time-varying DARM model parameters.

\section{Limits on Estimation of Systematic Error on Calibrated Strain}
\label{app:calibsyserrorlimits}

    \Sref{sec:syserrbackgnd} introduced historical and current methods used to estimate the systematic error on the detector response $\eta_R$ (see \eref{eq:etaR}).
    The primary limit on the instantaneous estimate of $\eta_{R}^{\textrm{(meas)}}$ has historically come from the uncertainty of the GPR fit\footnote{The superscript (meas) refers to the measurement of a given quantity.}.
    We can measure the error, $\eta_{R;A_{i}}^{\textrm{(meas)}}$ and $\eta_{R;C}^{\textrm{(meas)}}$, at any given time with the accuracy and precision limited by the absolute reference system and our knowledge of the well-known parameters. 
    It is the model of those errors that has high uncertainty, for several reasons.
    First, the model is informed by many repeated measurements that are taken outside of the time in question. 
    Second, those measurements are conditioned before fitting with well-known components (e.g. electronics) divided out.
    Third, the GPR is made with an idealistic kernel of hyperparameters that assumes little-to-no frequency-dependent systematic error, rather than the reality of a multitude of frequency-dependent features. 
    Fourth, we tune the bounds of the hyperparameters to cover all systematic error possibilities within the epoch, resulting in an uncertainty that is conservatively high.
    Finally, the GPR fit is propagated through the loop topology of the response function, which in the most sensitive region of the detector \emph{amplifies} any errors that may have been in the original GPR fits.
    For these reasons, the claimed uncertainty in $\eta_{R}$ for any instant in time, let alone for the entire epoch, is likely several factors larger than the underlying true systematic error at any given time.
    Figure 15 in \cite{sun2020characterization} shows this well; the uncertainty at each measured frequency point is barely visible on the scale, where the modeled systematic error is an appreciable portion of the plot.

As described in section 2.1, 2.2, and 2.3 of \cite{sun2020characterization}, some of the components involved in the systematic error estimation include the signal processing electronics and digital portions of the DARM loop. 
These components are exactly the portions that are relatively easy and inexpensive to change, and thus they've been changed frequently to improve detector sensitivity.
As the detection rate is proportional to the sensitivity cubed, even the most minor change in detector sensitivity will out-weigh the desire to keep things static -- allowing for a stable calibration and fixed understanding of the systematic error thereof. 
Every time a change to either $A$ or $C$ is made, it necessarily ``resets the clock'' on the model of the systematic error (c.f. Figure 1 in \cite{sun2020characterization}): updates to the well-known corners of the model parameter set are made, new reference-time measurements of $A$ and $C$ are taken to obtain ``core'' parameters, $A^{\textrm{(model)}}$ and $C^{\textrm{(model)}}$ and the filters to produce $h_{\textrm{det}}^{\textrm{(model)}}$ are updated, and $\eta_{R;A_{i}}^{\textrm{(meas)}}$ and $\eta_{R;C}^{\textrm{(meas)}}$ are reassessed.
Only the precision of the fit to $\eta_{R;A_{i}}^{\textrm{(meas)}}$ and $\eta_{R;C}^{\textrm{(meas)}}$ improves with time as repeat measurements are collected and added to the fit.
Until the detectors response is changed, and a new epoch begins.

\medskip

\printbibliography

\end{document}